\documentclass[aps,prl,twocolumn,groupedaddress,longbibliography]{revtex4-2}
\usepackage{graphicx}
\usepackage{dcolumn}
\usepackage{bm}
\usepackage{amsfonts}
\usepackage{xspace}
\usepackage{color}
\usepackage{epstopdf}
\usepackage{multirow}
\usepackage{amsmath}
\usepackage{float}
\usepackage[colorlinks=true, urlcolor=blue, linkcolor=blue, citecolor=blue]{hyperref}
\usepackage{cleveref}
\usepackage{amsmath}
\usepackage{mathtools}
\usepackage{enumitem}
\usepackage[section]{placeins}
\usepackage{soul}

\begin{document}

\preprint{}

\title{Stochastic transport of a Goldstone mode in a self-organized atomic crystal}

\author{Zhanhai Yu}
\thanks{These authors contributed equally to this work.}
\affiliation{Graduate School of China Academy of Engineering Physics, Beijing 100193, China}

\author{Di Xiang}
\thanks{These authors contributed equally to this work.}
\affiliation{Graduate School of China Academy of Engineering Physics, Beijing 100193, China}

\author{Xiaotian Zhang}
\thanks{These authors contributed equally to this work.}
\affiliation{Graduate School of China Academy of Engineering Physics, Beijing 100193, China}

\author{Hao Zhang}
\email[hzhang@gscaep.ac.cn]{}

\affiliation{Graduate School of China Academy of Engineering Physics, Beijing 100193, China}

\date{\today}

\begin{abstract}

Spontaneous breaking of a continuous symmetry produces a massless Goldstone mode that can evolve across a degenerate manifold at zero energy cost. Goldstone modes have been identified primarily through excitation spectra, mode softening or collective oscillations. However, their time-domain transport under intrinsic fluctuations and dissipation has remained largely unexplored. Here we directly track the stochastic transport of a Goldstone mode in a self-organized atomic crystal inside an optical ring cavity. The ring cavity maps the order-parameter phase onto the real-space position of the emergent crystal. Without any external perturbation, fundamental photon-scattering recoil drives the collective transport, while cavity dissipation generates friction. We monitor individual trajectories of the atoms and their self-generated optical lattice by measuring the cavity output phase.  We find that the diffusion constant decreases as $1/N$, indicating that all atoms move collectively as a rigid object rather than independently. By tuning the Langevin driving force and cavity-mediated damping, we show that the normalized diffusion constant collapses onto a single universal curve. This work extends the study of continuous symmetry breaking from excitation-frequency measurements to real-time tracking of transport, and opens routes for studying non-equilibrium collective transport, phonon dynamics, and defect formation in driven-dissipative quantum matter.

\end{abstract}

\maketitle

\begin{figure*}[tb]
\centering
\includegraphics[width=\textwidth]{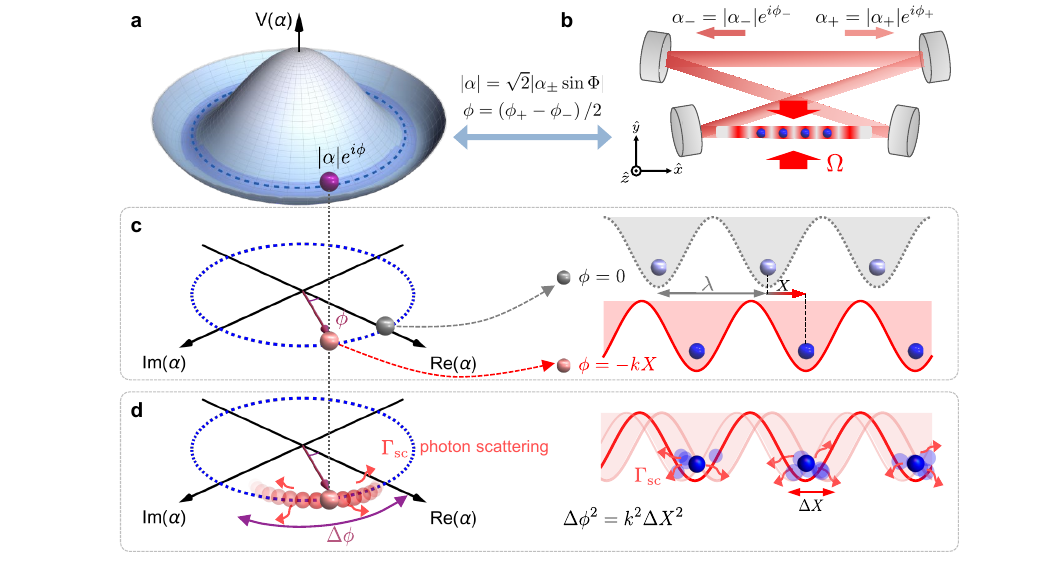}
\caption{\label{fig:1} \textbf{Ring-cavity system for Goldstone-mode dynamics.}
\textbf{a}, Effective Mexican-hat potential for the atom-cavity system with the $U(1)$ symmetry. The purple sphere denotes a symmetry-broken collective state of the atoms and their self-generated optical lattice, with the order-parameter amplitude $|\alpha|$ and azimuthal phase $\phi$.
\textbf{b}, Experimental setup. Inside the ring cavity, atoms driven by a transverse standing-wave pump field with Rabi frequency $\Omega$ emit light into two counter-propagating traveling-wave cavity modes $\hat a_\pm$, with complex amplitudes $\alpha_\pm=|\alpha_\pm|e^{i\phi_\pm}$. The order parameter and cavity fields have the relations $|\alpha|=\sqrt{2}|\alpha_\pm \sin \Phi|$ and $\phi=(\phi_+-\phi_-)/2$, where $\Phi$ is a common-mode phase.
\textbf{c}, Direct mapping from the $U(1)$ phase space onto the collective state in real space. Above the superradiance phase transition threshold, atoms self-organize into, and are trapped by an optical lattice that they themselves generate. The lattice emerges at a spontaneously selected position, breaking the continuous $U(1)$ symmetry. The gray point denotes a state with $\phi=0$, corresponding to the gray dashed lattice, whereas the red point denotes a state with $\phi=-kX$, corresponding to the red lattice translated by $X$. Thus, the phase $\phi$ is directly related to the lattice position $X$, and the amplitude $|\alpha|$ is related to the emergent lattice depth and the atomic structure.
\textbf{d}, Stochastic transport of the spatially uniform Goldstone mode. Intrinsic photon-scattering events at rate $\Gamma_{\mathrm{sc}}$ imparts stochastic recoil to the collective object of atoms and the optical lattice, driving diffusion of the collective center-of-mass coordinate $X$ and producing a displacement $\Delta X$. Through the phase--position relation, the corresponding variances are related by $\Delta\phi^2=k^2\Delta X^2$.
}
\end{figure*}

Spontaneous breaking of a continuous symmetry is a cornerstone of modern physics, underlying phenomena from magnetism and superfluidity to particle physics. Its defining consequence is the emergence of a massless Goldstone mode \cite{nambu1961dynamical,goldstone1961field,goldstone1962broken} — a collective excitation that wanders through a continuous manifold of degenerate states at zero energy cost. This mode has been pursued across diverse experimental platforms through compressional oscillations in harmonically trapped quantum gases \cite{natale2019excitation,guo2019lowenergy,tanzi2019supersolid,norcia2021twodimensional,chisholm2026probinga}, spectral narrowing in the optical response of microcavity polariton condensates \cite{claude2025observation}, and softening of the excitation gap in quantum gases in optical lattices \cite{endres2012higgs} and crossed optical cavities \cite{leonard2017supersolidd,leonard2017monitoringb}. In each case, observations have focused on detecting a decrease in excitation energy — mode softening — which signals that the excitation is approaching the gapless regime. The outstanding challenge is not only to identify a soft mode, but to directly follow the Goldstone coordinate itself: to watch the collective motion as it evolves from one state to another, covering the full continuous space of degenerate states. Here we address the question: what drives the transport of the collective ordered state, selected by continuous symmetry breaking, when it is left to evolve under its own intrinsic fluctuations and dissipation?

We realize a self-organized atomic crystal in an optical ring cavity and track the real-space motion of its Goldstone coordinate. Above the superradiant phase transition, atoms collectively scatter light into the cavity and self-organize into a crystalline structure, self-trapped by the dynamical optical lattice they themselves generate \cite{gopalakrishnan2009emergentb,gopalakrishnan2010atomlight,ritsch2013coldb,mivehvar2021cavity}. Discrete Z$_2$ symmetry breaking has been observed in atoms coupled to Fabry-Pérot cavities \cite{black2003observationa,baumann2011exploringb,ho2025optomechanical}, whereas continuous $U(1)$ symmetry breaking has been realized in two crossed Fabry-Pérot cavities \cite{leonard2017supersolidd,leonard2017monitoringb}. In contrast, the traveling-wave geometry of an optical ring cavity guarantees $U(1)$ symmetry by construction (Fig.~\ref{fig:1}a). Inside the ring cavity, two counterpropagating traveling-wave cavity modes are populated by light scattered from the atoms (Fig.~\ref{fig:1}b). Together with the pump field, these modes generate a dynamical optical lattice in which the atoms are trapped. The ring cavity maps the phase of the $U(1)$ symmetry-broken order parameter onto the real-space position of the atomic crystal trapped by their self-generated optical lattice (Fig.~\ref{fig:1}c) \cite{mivehvar2018drivendissipativeb}. By heterodyne detection of the transmitted cavity field phase, we monitor in real time the individual trajectories of this collective motion.

The ring cavity provides a translationally invariant real space, so the self-organized atomic crystal can undergo stochastic transport along the ring cavity axis, without any external force, covering the entire $U(1)$ phase space from 0 to $2\pi$ (Fig.~\ref{fig:1}d). The driving force for the Goldstone mode does not come from an external perturbation, but from the intrinsic fundamental off-resonant photon-scattering of the superradiant lattice itself. As the atomic crystal slides in space, it experiences a damping force resulting from the finite cavity response, establishing a connection between symmetry-breaking collective motion and cavity optomechanics. We observe that the atomic crystal undergoes diffusive motion. By tuning the Langevin driving force (via the photon-scattering rate $\Gamma_{\mathrm{sc}}$) and the damping (via the pump–cavity detuning $\delta$), we track how the diffusion constant depends on the microscopic parameters. When normalized by the lattice spacing and the atomic scattering rate, the measured diffusion collapses onto a universal curve governed solely by the atom number $N$ and a normalized damping coefficient $\tilde{\beta}$. We thereby obtain a controllable open non-equilibrium setting in which to measure the stochastic transport of the Goldstone mode.

Notably, the measured diffusion constant follows the scaling $\mathcal{D} \propto 1/N$ with atom number — demonstrating that all atoms move as a single rigid object rather than as independent particles. Unlike conventional Brownian motion, where particles diffuse independently in a passive static bath, our system features atoms strongly coupled to an active lattice that is dynamically generated by the atoms themselves and back-acts on their motion. It is this cavity-mediated feedback that locks the atoms together. This collective rigidity, imprinted on a single observable — the cavity phase — transforms the Goldstone mode from a spectroscopic feature into a controllable mechanical degree of freedom.

\subsection*{Mapping the Goldstone mode onto real space}
\begin{figure*}[ht]
\centering
\includegraphics[width=\textwidth]{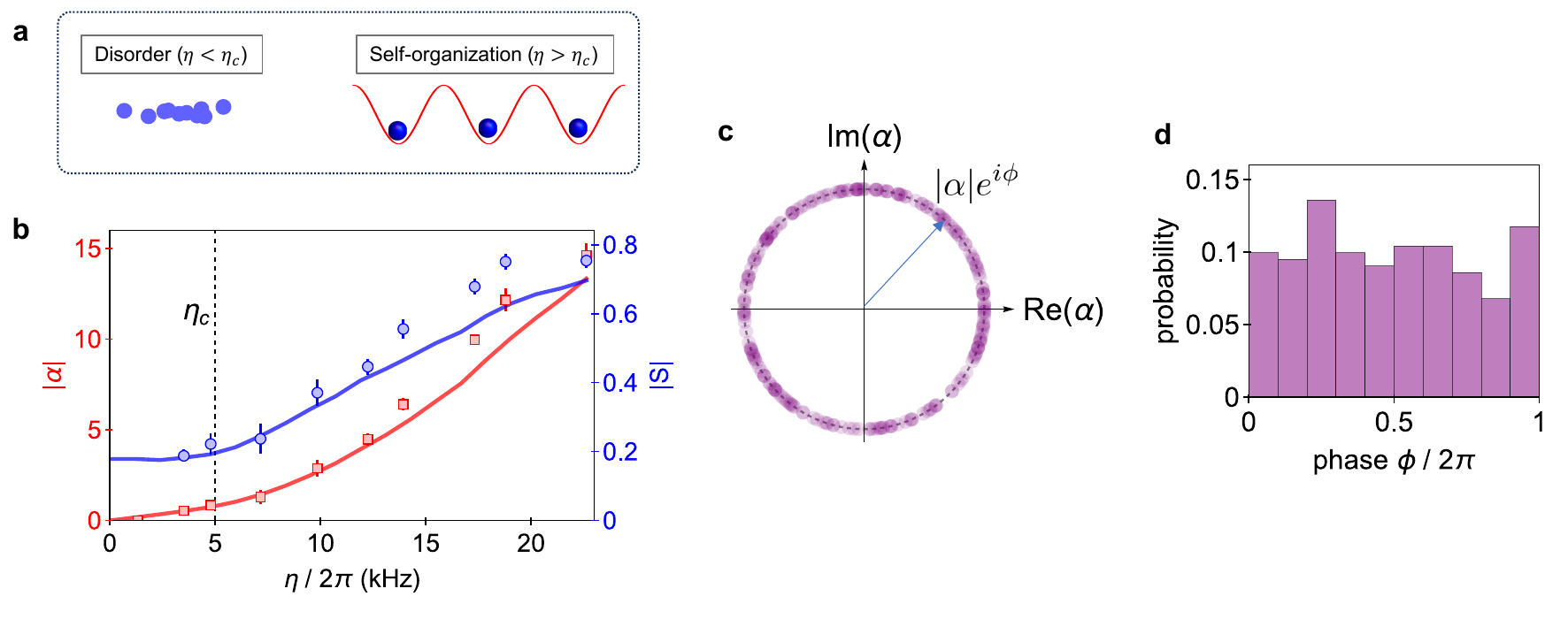}
\caption{\label{fig:2} \textbf{Superradiant self-organization and spontaneous $U(1)$ symmetry breaking.} \textbf{a}, Illustration of the self-organization. Below the threshold ($\eta<\eta_c$), atoms remain disordered and do not generate an optical lattice. Above threshold ($\eta>\eta_c$), atoms self-organize into an emergent lattice with period $\lambda$. \textbf{b}, Order-parameter amplitude $|\alpha|$ and atomic structure factor $|S|=\left|\frac{1}{N}\sum_{j=1}^{N}e^{ikx_j}\right|$
versus effective pump strength $\eta=g\Omega/(2\Delta)$. The vertical dashed line marks the threshold $\eta_c$. The non-zero value of $|S|$ at zero pump strength arises from finite-size effects. Solid lines are parameter-free theoretical calculations. Data are shown for $N=20$, $\delta = -2\pi\times 30$ kHz. Error bars denote the standard error of the mean.
\textbf{c}, Distribution of the order-parameter phase $\phi$ in the complex plane measured in 221 independent realizations above threshold, prepared from randomized initial atomic configurations. 
\textbf{d}, Histogram of the phase $\phi$, showing an approximate uniform distribution over $[0,2\pi)$, consistent with spontaneous selection of the $U(1)$ symmetry-broken phase. Data in \textbf{c,d} are taken at $N = 10$, $\delta = -2\pi\times 50$ kHz, $\Omega = 2\pi\times 125$ MHz.}
\end{figure*}
We realize a cavity-mediated many-body system by coupling a $^{87}$Rb atomic array to an optical ring cavity (Fig.~\ref{fig:1}b, and Methods) \cite{zhang2024cavity}. Atoms interact with two counter-propagating cavity traveling-wave modes $\hat{a}_\pm$ and are driven by a $\sigma^-$-polarized transverse standing-wave pump field of Rabi frequency $\Omega$. In the dispersive regime, adiabatic elimination of the internal atomic states yields the effective Hamiltonian
\begin{equation} \label{eqs:Ham0}
\begin{aligned}
\frac{\hat{H}}{\hbar} &=-\delta_s\left(\hat{a}_+^\dag\hat{a}_++\hat{a}_-^\dag\hat{a}_-\right)
+\frac{g^2}{\Delta}\left(\hat{a}_+\hat{a}_-^\dag\sum_{j=1}^Ne^{2ikx_j}+h.c.\right)
 \\ &-\eta\left(i\hat{a}_+\sum_{j=1}^Ne^{ikx_j}+i\hat{a}_-\sum_{j=1}^Ne^{-ikx_j}+h.c.\right),
\end{aligned}
\end{equation}
where $g=2\pi\times0.7$ MHz is the effective cavity vacuum Rabi frequency, $\Delta=-2\pi\times2$ GHz is the pump-atom detuning, $N$ is the total atom number, $x_j$ is the position of the $j$-th atom, $k = 2\pi/\lambda$ is the cavity wavevector of wavelength $\lambda=780$ nm. The dispersive atomic shift modifies the effective pump–cavity detuning to $\delta_s=\delta-Ng^2/\Delta$, where $\delta$ is the bare pump-cavity detuning. The cavity modes are driven by pump light coherently scattered by the atoms, with effective coupling $\eta=g\Omega/(2\Delta)$. The Hamiltonian in Eq.~(\ref{eqs:Ham0}) possesses a $U(1)$ symmetry. It is invariant under the simultaneous translation of all atoms $x_j\to x_j+\delta x$, accompanied by the transformation of the cavity fields $\hat{a}_{\pm}\to e^{\mp ik\delta x}\hat{a}_{\pm}$ \cite{mivehvar2018drivendissipativeb}. 

When atoms are driven by the transverse pump, they scatter photons into the cavity and build up cavity fields. For atoms distributed with the structure factor $S=\frac{1}{N}\sum_{j=1}^{N} e^{ikx_j}$, the mean cavity-mode amplitudes $\hat{a}_\pm$ are given by $\alpha_+=|\alpha_+|e^{i\phi_+}=N\eta S^*/(-i\delta_s+\kappa/2)$ and $\alpha_-=|\alpha_-|e^{i\phi_-}=N\eta S/(-i\delta_s+\kappa/2)$, neglecting the mutual coupling between the two cavity modes, where $\kappa/(2\pi)=34$ kHz is the cavity linewidth. Their phases $\phi_\pm=\Phi\mp\arg{(S)}$ are set by the atomic structure factor $S$ and a common-mode phase $\Phi=\arctan(2\delta_s/\kappa)$, where $\Phi$ is independent of atomic positions and is set solely by the cavity dispersive response. These cavity fields, interfering with the transverse pump field, generate an emergent lattice for the atoms, $U(x)=-\hbar\eta(i\alpha_+-i\alpha_-^*)e^{ikx}+h.c. = -U\cos{\left(kx+\phi\right)}$, where $\phi = (\phi_+ - \phi_-)/2=-\arg{(S)}$ is determined by the atomic positions through the structure factor $S$ and, in turn, sets the lattice position. Thus, the atoms generate an optical lattice whose position follows their collective configuration. The lattice depth is $U=-4\hbar\eta|\alpha_\pm|\sin{\Phi}$. Note that the common-mode phase $\Phi$ affects only the lattice depth, not its position. This lattice can trap atoms with spacing $\lambda$, thereby sustaining both the cavity fields and the atomic ordering. The lattice depth $U$ can be equivalently written as $U=-2\sqrt{2}\hbar\eta|\alpha|$, where we define the complex order parameter as $\alpha=\frac{1}{\sqrt{2}}(-i\alpha_++i\alpha_-^*)$. The order parameter $\alpha$ has a clear physical interpretation: the amplitude $|\alpha|$ is proportional to the emergent lattice depth and the atomic structure factor, whereas its phase $\phi$ determines the lattice position.

\begin{figure*}[htbp]
\includegraphics[width=\textwidth,angle=0]{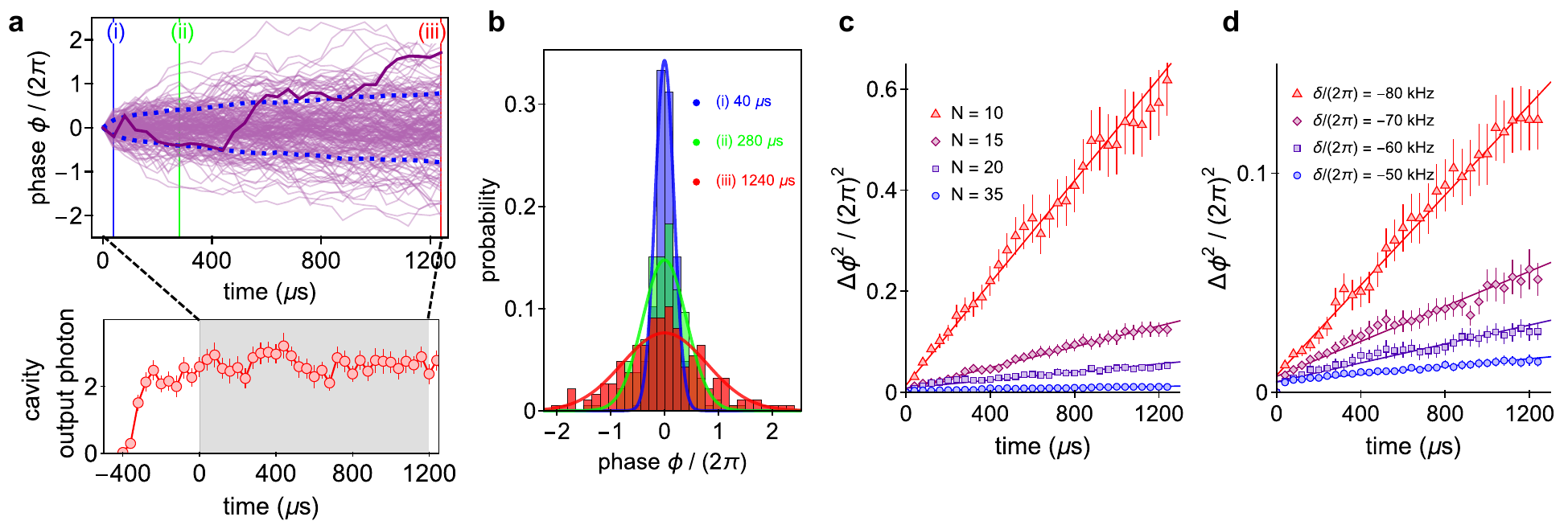}
\caption{\label{fig:3} \textbf{Real-time observation of stochastic Goldstone-mode transport.} \textbf{a}, The phase $\phi$ undergoes stochastic motion over time (top) while the cavity photon number remains approximately constant after a settling time (bottom). The transmitted cavity field is measured by heterodyne detection. The panel shows 186 experimental trajectories. The dark purple line is one representative trajectory. Blue dashed lines indicate standard deviation $\pm\Delta\phi(t)$. Data are taken at $N=10$, $\delta=-2\pi\times 80$ kHz, and $\Omega=2\pi\times125$ MHz. \textbf{b}, Phase distributions at the times marked in \textbf{a}. Solid lines are Gaussian fits. \textbf{c}, Phase variance $\Delta\phi^2$ versus time for different atom numbers $N$ at fixed pump Rabi frequency $\Omega = 2\pi\times 125$ MHz and pump–cavity detuning $\delta = -2\pi\times 80$ kHz. \textbf{d}, Phase variance versus time for different pump–cavity detunings $\delta$ at fixed atom number $N=15$ and pump Rabi frequency $\Omega = 2\pi\times 125$ MHz. All solid lines are linear fits from which we extract the diffusion constant $\mathcal{D}$. Error bars denote the standard error of the mean.}
\end{figure*}

Solving the coupled equations of motion for the cavity fields and atoms, we find that for atoms at finite temperature and near the phase transition, the effective potential takes the form $V(\alpha)=-A|\alpha|^2+B|\alpha|^4$  (see Methods for details). This has the characteristic Mexican-hat form shown in Fig.~\ref{fig:1}a, reflecting the continuous degeneracy associated with the $U(1)$ manifold. The microscopic origin of the transition is contained in the coefficient $A$. As the pump strength is increased to reach a critical value $\eta_c$ when the potential energy gained from the self-organization overcomes the thermal energy, the coefficient $A$ for the potential quadratic term changes sign, destabilizing the disordered state with $|\alpha|=0$ in favor of spontaneous order with non-zero $|\alpha|$. Above this threshold, the atoms self-organize to a $\lambda$-periodic crystal, and the emergent cavity lattice acquire non-zero amplitudes $|\alpha|$. The phase of the order parameter selects one value $\phi \in [0, 2\pi)$, reflecting the breaking of the $U(1)$ symmetry, as illustrated in Fig.~\ref{fig:1}c. In the regime $k_\text{B} T \ll U$ where the atomic temperature $T$ is much lower than the lattice depth $U$, all atoms are tightly confined, forming a rigid crystalline structure. In this limit, the phase of the cavity lattice directly tracks the collective center-of-mass motion, $\phi = -kX$, where $X = \frac{1}{N}\sum_j x_j$ is the center of mass coordinate of the atoms. 

\subsection*{$U(1)$ symmetry breaking}
We first characterize the $U(1)$-breaking superradiant transition experimentally. Atoms are loaded into the ring cavity in random tweezer configurations, and the pump power is ramped up while the tweezer traps are adiabatically switched off. We monitor the cavity field as a function of pump strength $\eta$. Below the critical pump strength $\eta_c$, the atoms remain disordered and the cavity field is near zero. Above $\eta_c$, the cavity amplitude $|\alpha|$ increases in agreement with our parameter-free theory (Fig.~\ref{fig:2}a,b). The structure factor $S$, extracted from the measured $|\alpha|$, simultaneously approaches unity, indicating self-organization into the $\lambda$-periodic lattice. Because the atoms are prepared with a random spatial configuration in each realization, the phase $\phi$ of the emergent lattice selected varies from shot to shot. The phase $\phi$ can be extracted by measuring the phase of the transmitted cavity field. Using heterodyne detection (see Methods for details), we directly observe the $U(1)$ symmetry breaking: the phase $\phi$ spans the full range from 0 to $2\pi$ (Fig.~\ref{fig:2}c,d).

\subsection*{Stochastic transport of the Goldstone mode}
Once the atomic crystal and optical lattice form inside the ring cavity, all translated lattice configurations are energetically equivalent. Even without applying an additional modulation or probe, the atoms and their self-generated optical lattice can translate along the ring cavity axis without a restoring force. We track the displacement of the atomic crystal by monitoring the time evolution of cavity transmitted phase. The phase $\phi$ exhibits stochastic motion while the cavity photon number remains nearly constant, indicating the atoms remain self-organized and move collectively with the optical lattice (Fig.~\ref{fig:3}a). We perform repeated measurements of many phase trajectories (Fig.~\ref{fig:3}a,b). The phase variance $\Delta\phi^2$, which reflects the collective diffusion of the self-organized atoms by $\Delta\phi^2= k^2 \Delta X^2$, grows linearly in time (Fig.~\ref{fig:3}c,d  ). Fitting the atomic crystal center-of-mass coordinate $\Delta X^2 = 2 \mathcal{D} t + c$, we extract the diffusion constant $\mathcal{D}$. Diffusion slows with increasing atom number $N$ and decreasing the pump–cavity detuning $\delta$. The two knobs — $N$ and $\delta$ — control, respectively, the collective mass of the diffusing object and the effective friction, allowing us to scan and test their microscopic contributions.

\begin{figure*}[t]
\includegraphics[width=\textwidth,angle=0]{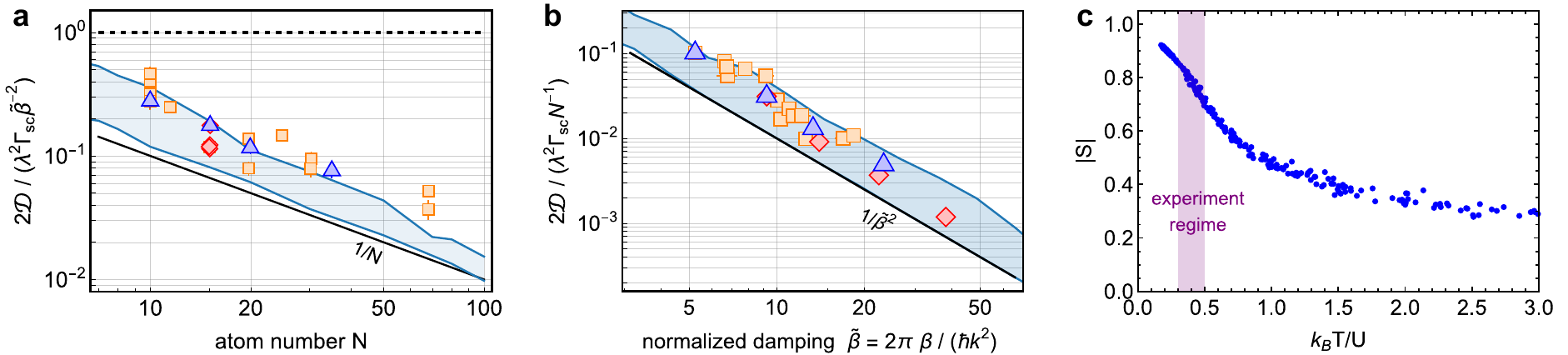}
\caption{\label{fig:4} \textbf{Universal scaling of the normalized diffusion constant.} \textbf{a}, Normalized diffusion constant versus atom number $N$. The solid line represents the zero-temperature theory, $\mathcal{D} \propto 1/N$. The dashed line at unity indicates the diffusion constant for independent atoms. \textbf{b}, Normalized diffusion constant versus the normalized damping coefficient $\tilde{\beta}$. The solid line is the zero-temperature theory, $\mathcal{D} \propto 1/\tilde{\beta}^2$. Small deviations of the measurements from the zero-temperature prediction arise from finite temperature effects. Blue triangles and red diamonds correspond to the data in Fig.~\ref{fig:3}d and Fig.~\ref{fig:3}e, respectively. Orange squares denote measurements taken under other conditions, including simultaneous scans of pump Rabi frequency $\Omega$, atom number $N$, and pump-cavity detuning $\delta$. Shaded bands are finite-temperature simulations at $k_B T/U=0.4(1)$. Error bars denote the standard error of the mean. \textbf{c}, Numerical simulation of the structure factor $|S|$ versus temperature. The structure factor decreases with increasing temperature, showing a crossover from a rigid self-organized atomic crystal melted to less correlated thermal motion. The purple shaded region marks the experimental regime.}
\end{figure*}

\subsection*{Universal scaling of the diffusion constant}
We now examine the microscopic origin of the diffusion phenomenon, which arises from two mechanisms in our system. First, stochastic off-resonant photon scattering imparts random recoil to each atom. The motion of any individual atom contributes to displacing the entire optical lattice, which binds and moves all atoms together. These stochastic forces thus drive the collective motion of both atoms and the lattice. Second, the cavity lattice lags behind atomic motion due to the finite response time of the cavity field $1/\kappa$. As the atoms move, the cavity field—collectively generated by atoms—adjusts to the new atomic configuration with a delay, giving rise to a velocity-dependent friction force. 

To quantify diffusion, we solve the coupled equations of motion for the cavity fields and atomic positions derived from Eq.~(\ref{eqs:Ham0}) above threshold. As atoms move, the cavity fields deviate from their steady-state values for a stationary atomic configuration. Expanding the cavity fields to first order in the collective velocity $\dot{X}$ as $\langle\hat{a}_\pm(t)\rangle=\alpha_\pm+\dot{X}\delta\alpha^{(1)}_\pm$  \cite{vuletic2000laserb}, where $\alpha_\pm$ are the steady-state fields, and $\delta\alpha^{(1)}_\pm$
are the first-order corrections arising when the atomic crystal is moving, we obtain a damping force $-\beta \dot{X}$ acting on each atom, proportional to $\dot{X}$, with $\beta$ the damping coefficient for a single atom (see Methods for details).
The collective coordinate $X$ then obeys the Langevin equation
\begin{equation} \label{eqs:EomX}
\begin{aligned}
M\ddot{X}=-N\beta\dot{X}+F(t),
\end{aligned}
\end{equation}
where $M=Nm$ is the total mass of atoms. The total stochastic force acting on the whole rigid body of atoms is $F(t) = \sum_{j=1}^N f_j(t)$, where $f_j(t)$ is the stochastic recoil force on the $j$-th atom arising from off-resonant photon scattering. The noise correlation strength of $F(t)$ that acts on the atomic crystal is $S_{FF} = N\hbar^2 k^2 \Gamma_{\mathrm{sc}}$, where $\Gamma_{\mathrm{sc}}$ is the photon-scattering rate of a single-atom and $\hbar k$ is the recoil momentum of a single-photon \cite{vuletic2001threedimensionala,hosseini2017cavitya}. The scattering rate can be written as $\Gamma_{\text{sc}}=\left(\frac{3}{10}+\frac{2C\kappa^2}{4\delta^2+\kappa^2}\right)\Gamma_{\text{fs}}$, where $C$ is the cavity cooperativity \cite{tanji-suzuki2011interactionb} and $\Gamma_{\text{fs}}$ is the free-space scattering rate of a single atom inside the cavity driven by the pump laser. The two terms correspond, respectively, to free-space scattering projected along the cavity axis and to scattering into the cavity mode. The friction damping coefficient is $\beta=k^2 U\frac{2\kappa}{4\delta^2+\kappa^2}\left(1+\frac{\delta}{\delta-2\frac{Ng^2}{\Delta}} \right)$, proportional to the cavity lattice depth $U$ and controlled by the cavity detuning $\delta$ and linewidth $\kappa$. The damping coefficient recovers the form of cavity cooling in the ring cavity \cite{ganglColdAtomsHighQ2000}. The diffusion constant is therefore given by the generalized Einstein relation — the ratio of stochastic noise to damping:
\begin{equation}\label{eqD}
\mathcal{D} = \frac{S_{FF} }{2(N\beta)^2}=\frac{N\hbar^2 k^2 \Gamma_{\text{sc}}}{2(N\beta)^2}.
\end{equation}

To reveal the universal behavior of the diffusion dynamics, we note that $\mathcal{D}$ has the dimensions of $\text{[length]}^2/\text{[time]}$. In our system, the natural length scale is the lattice wavelength $\lambda$, and the natural time scale is set by the inverse of the scattering rate $\Gamma_{\text{sc}}^{-1}$. Therefore, we normalize $\mathcal{D}$ in Eq.~(\ref{eqD}) as
\begin{equation}\label{normD}
  \frac{2 \mathcal{D}} {\lambda^2 \Gamma_{\text{sc}}}=\frac{1}{N\tilde{\beta}^2},
\end{equation}
where $\tilde{\beta} = 2\pi\beta/(\hbar k^2)$ is the normalized damping coefficient. We scan the atom number $N$, pump-cavity detuning $\delta$, and pump Rabi frequency $\Omega$, and measure the collective diffusion constant $\mathcal{D}$. In Fig.~\ref{fig:4}a, we vary $N$ over nearly one order of magnitude and plot the normalized $2 \mathcal{D}/(\lambda^2 \Gamma_{\text{sc}})$ as a function of $N$. Compared to the case of independent atoms where the diffusion constant is independent of $N$, our collective diffusion slows down by a factor of $\sim25$ at $N=70$. This collective slowing is reminiscent of collective recoil sharing in the Mössbauer effect \cite{mssbauer1958kernresonanzfluoreszenz,eyges1965physics}, in which a heavy composite object is less susceptible to recoil. In our system, however, the recoil momentum originates from the intrinsic fundamental stochastic photon scattering rather than an externally applied mechanical force, and the recoil is shared by the entire self-organized atomic crystal and their self-generated optical lattice. In Fig.~\ref{fig:4}b, we plot the normalized diffusion constant $2 \mathcal{D}/(\lambda^2 \Gamma_{\text{sc}})$ as a function of the normalized damping coefficient $\tilde{\beta}$. Here, the physical meaning of $\tilde{\beta}$ is the cavity-mediated damping in units of $\hbar k^2$. The measured diffusion constants collapse onto a single curve, demonstrating a universal scaling of collective transport in an open, driven-dissipative many-body system.

The measured diffusion follows the predicted scaling $\mathcal{D} \propto 1/( N \tilde{\beta}^2)$ with small deviations from the zero-temperature theory. We attribute these deviations to finite-temperature effects at nonzero $k_B T/U$. By independently measuring the temperature and the optical-lattice depth, we estimate $k_B T/U=0.4(1)$ in our experiment. To quantify the corresponding corrections, we numerically calculate the diffusion constant across the experimental temperature range (see Methods). The resulting range of the calculated diffusion constant is shown by the shaded bands in Fig.~\ref{fig:4}a,b. Thermal fluctuations introduce additional atomic position fluctuations, leading to deviations from the ideal zero-temperature behavior. To assess thermal effects on atomic order, we numerically simulate the structure $|S|$ at different temperatures as shown in Fig.~\ref{fig:4}c. The structure decreases with increasing temperature, indicating thermal melting of the self-organized lattice. At low temperature, atoms are trapped by their self-generated optical lattice and move collectively with it as an approximately rigid body. At high temperature, thermal fluctuations weaken the atomic order and lead to less correlated motion. At the experimentally measured temperature, the atoms remain well organized in the self-generated lattice.

\subsection*{Discussion and outlook}
While the present observation focuses on a Goldstone mode within a single-wavelength lattice—where the system is mechanically stiff—future extensions are readily accessible. By driving multiple cavity modes \cite{guo2021opticala,kroeze2025directly,marsh2025multimode}, we can excite lattice vibrations and tune the rigid crystal into a programmable phonon medium. By quenching across the superradiant transition, we can induce phase kinks and domain formation as different regions select distinct phases, making this platform a testbed for studying Kibble‑Zurek dynamics \cite{zurek1985cosmological,klinder2015dynamicala,suzuki2025deconstructing}, enabling the real-time tracking of topological defect formation during spontaneous symmetry breaking \cite{pyka2013topological,yang2026topological}. Moreover, deliberately breaking the U(1) symmetry with external optical tweezers of programmable geometry  \cite{keesling2019quantum,chen2023continuous} will deform the degenerate manifold into engineered, non-degenerate energy landscapes, with relevance to questions of pinning \cite{haller2010pinning,keller2022selfpinning} and glassiness \cite{gopalakrishnan2011frustration,kroeze2025directly,marsh2025multimode} in correlated quantum matter. More broadly, by promoting a Goldstone mode from a spectral feature to a directly tracked, controllable degree of freedom, our results establish an experimental handle that bridges spontaneous symmetry breaking concepts from high‑energy and condensed‑matter physics to driven‑dissipative many-body dynamics.

\textbf{Acknowledgements} This work was supported by the National Key Research and Development Program of China (Grant No. 2022YFA1405302) and the National Natural Science Foundation of China (Grants Nos. 92565202, 12504425, and U2330401).

\setcounter{figure}{0}

\renewcommand{\figurename}{Fig.}
\renewcommand{\thefigure}{S\arabic{figure}}

\renewcommand{\theHfigure}{ED\arabic{figure}}
\setcounter{equation}{0}
\renewcommand{\theequation}{S\arabic{equation}}
\section{Methods}

\subsection{Experimental sequence}
$^{87}$Rb atoms are loaded into the ring cavity using an optical tweezer array that is positioned along the cavity axis, as described in Ref.~\cite{zhang2024cavity}. The tweezer array is generated by focusing 850-nm light through an aspheric lens ($\text{NA} = 0.5$). The array geometry is controlled using an acousto-optic deflector (AOD), with a nominal position resolution of approximately 5 nm set by the RF-frequency resolution of the arbitrary waveform generator. Atoms are first loaded into the tweezers and cooled by polarization-gradient cooling to about 30 $\mu$K in traps with a typical depth of 1.4 mK. To maximize coupling to the cavity, we optically pump the atoms into the $|F=2,~m_F=-2\rangle$ state with a $\sigma_-$-polarized laser resonant with the $|F=2\rangle \to |F'=2\rangle$ transition, together with a repumper beam resonant with the $|F=1\rangle \to |F'=2\rangle$ transition. 

\begin{figure*}[htbp]
\includegraphics[width=\textwidth,angle=0]{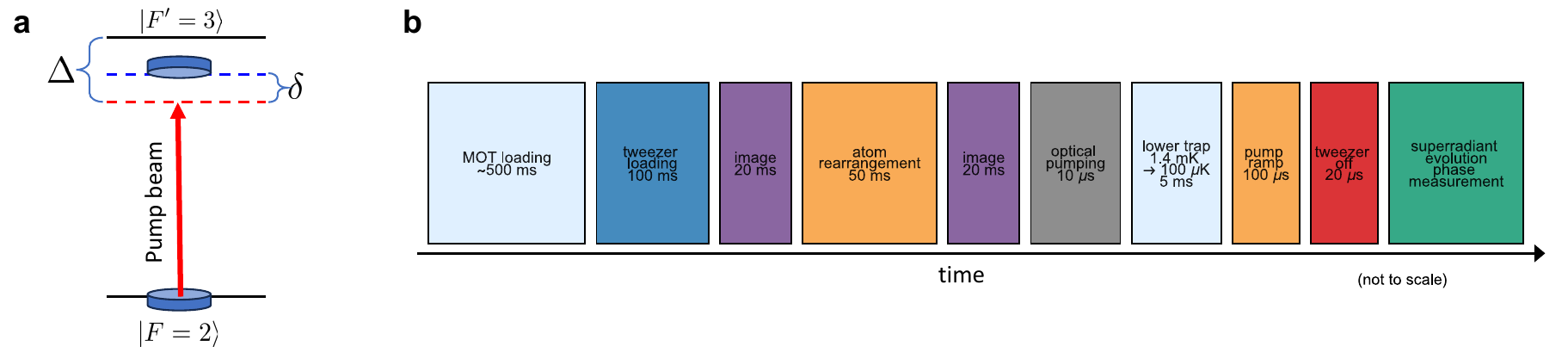}
\caption{\label{fig:S1}\textbf{Experimental sequence.}
\textbf{a,} Pump configuration for driving the superradiant transition. The pump beam couples the relevant hyperfine states with the pump--atom detuning $\Delta$. The cavity resonance is stabilized at a frequency detuned from the pump by $\delta$.
\textbf{b,} Experimental timing sequence. After the magneto-optical trap (MOT) loading, atoms are loaded into optical tweezers and rearranged into the desired configurations. The pump is ramped up over $100~\mu\mathrm{s}$, after which the optical tweezers are ramped to zero over $20~\mu\mathrm{s}$. For pump strengths above threshold, the system subsequently evolves into a self-organized superradiant state.}
\end{figure*}

The pump light that drives atoms to scatter the light into the cavity is a $\sigma^{-}$-polarized standing wave. It is formed by two counter-propagating beams along the $y$ direction with a beam waist of approximately $675~\mu\mathrm{m}$ and a pump-atom detuning $\Delta=-2\pi\times2~\mathrm{GHz}$ relative to the $F=2\rightarrow F'=3$ transition, as illustrated in Fig.~\ref{fig:S1}a.

During the experiment, we adiabatically transfer atoms from the initial tweezer array to the final cavity optical lattice that the atoms collectively generate. Before applying the pump laser that drives the atoms, the tweezer depth is adiabatically reduced from $1.4~\mathrm{mK}$ to approximately $100~\mu\mathrm{K}$. The pump laser intensity is then ramped to its target value over $100~\mu\mathrm{s}$ to drive atom-to-cavity scattering, while the repumper beam remains on to maintain the atoms in the $|F=2,~m_F=-2\rangle$ state. Once the pump laser reaches its target intensity, the optical tweezers are finally ramped to zero over $20~\mu\mathrm{s}$, as shown in Fig.~\ref{fig:S1}b. When the pump laser power exceeds the self-organization threshold, the atoms become trapped by the emergent cavity optical lattice. Removing the tweezer potential eliminates external pinning, allowing the atom-lattice system to translate freely along the cavity axis.

\subsection{Cavity parameters}
The cavity has a finesse of $\mathcal{F}=4.4(1)\times10^4$, a mode waist of $w_0\simeq7~\mu\mathrm{m}$, and an intensity decay rate of $\kappa=2\pi\times34~\mathrm{kHz}$ for both traveling-wave modes. Cavity birefringence produces two orthogonal linearly polarized modes separated by 1.76 MHz. We measure a single-atom cooperativity of $C\simeq12.5$ for each polarization mode, where $C=4g^2/(\gamma_a\kappa)$ and $\gamma_a=2\pi\times6.06~\mathrm{MHz}$ is the atomic natural linewidth. These values correspond to an atom–cavity coupling strength of $g\simeq2\pi\times0.80~\mathrm{MHz}$. In this work, we couple one or two rows of atoms to the cavity mode. Spatial averaging over the Gaussian cavity-mode profile reduces the effective cooperativity to $C_{\mathrm{eff}}\simeq10$, corresponding to an effective atom–cavity coupling strength of $g_{\mathrm{eff}}\simeq2\pi\times0.7~\mathrm{MHz}$. We use these effective values throughout the calibration and numerical simulations. The transmitted fields from the two traveling-wave modes are directed onto single-photon-counting modules with a total detection efficiency of about $10\%$.

\subsection{Phase measurement}
For a self-organized atomic lattice at low temperature, the atomic position can be written as $x_j\simeq X+m_j\lambda$, thus the structure factor depends only on the center of mass $S=e^{ikX}$. The two counter-propagating cavity modes acquire opposite position-dependent phases, with $\phi_+=\Phi-\arg{(S)}=\Phi-kX$ and $\phi_-=\Phi + \arg{(S)}=\Phi+kX$, where $\Phi$ is the cavity-response phase that is independent of atomic positions. The phase of the order parameter, which is also the phase in the emergent lattice, is therefore $\phi=(\phi_+-\phi_-)/2=-kX$. These relations show that the phase of a single cavity mode is therefore sufficient to determine the collective phase $\phi$ after removing a common-mode phase.

\begin{figure*}[htbp]
\includegraphics[width=\textwidth,angle=0]{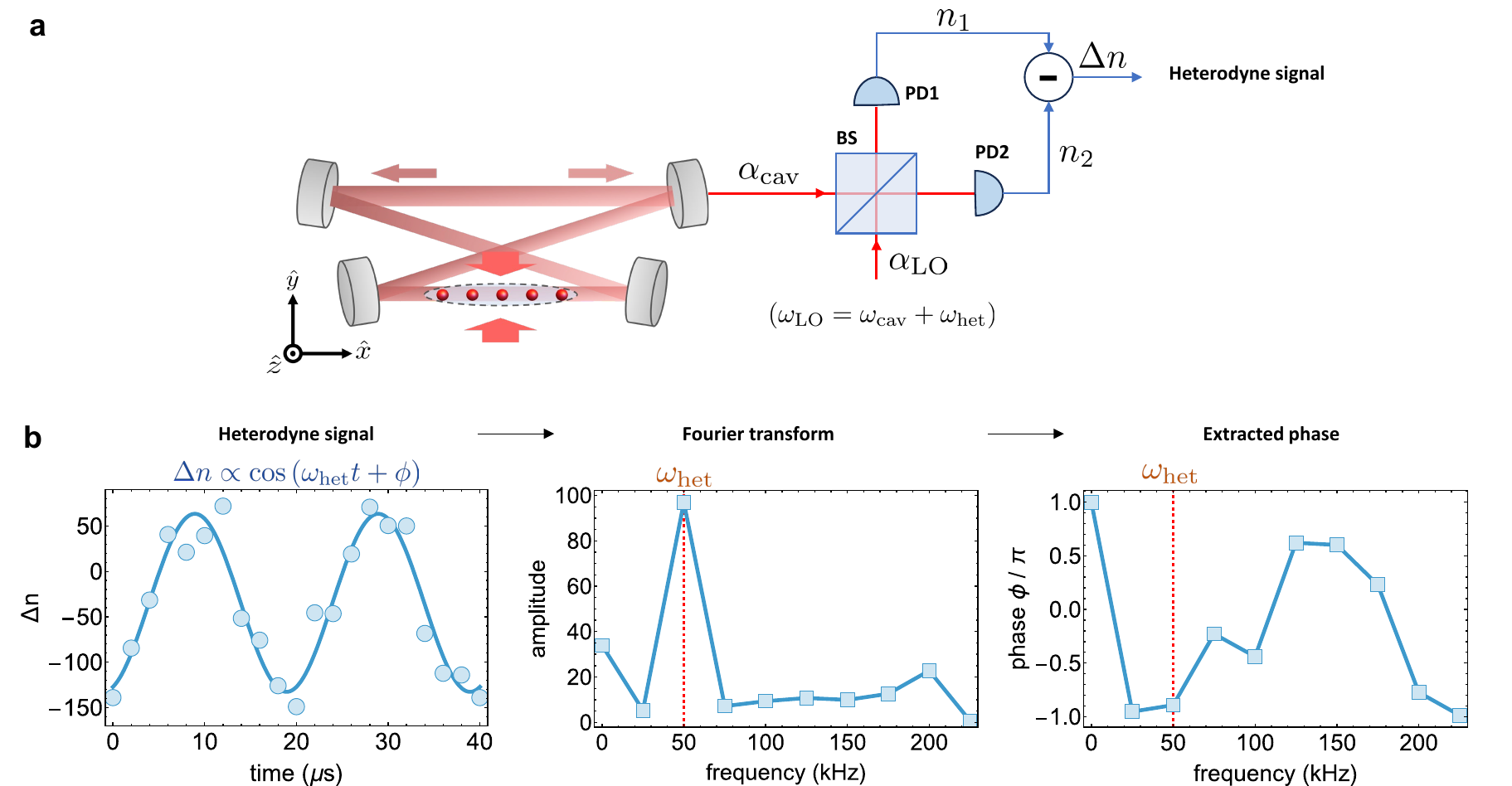}
\caption{\label{fig:S2}\textbf{Heterodyne phase detection.} \textbf{a,} Heterodyne detection of the cavity phase. The transmitted cavity field $\alpha_\text{cav}$ is mixed with a local oscillator $\alpha_\text{LO}$  on a beam splitter (BS), and the two output ports are detected by single-photon counting modules, PD1 and PD2. \textbf{b,} The differential photon-count signal gives a heterodyne beat signal. For each 40-$\mu$s window, the signal is Fourier transformed and the cavity phase is extracted from the complex Fourier component at the heterodyne frequency. A representative beat signal is shown together with its amplitude and phase spectra; red dashed lines mark the heterodyne frequency used for phase extraction.}
\end{figure*}

The cavity phase is measured using a heterodyne detection scheme, as illustrated in Fig.~\ref{fig:S2}a. The transmitted cavity field of $\hat{a}_+$ mode is mixed with a weak local oscillator (LO) beam detuned by $50~\mathrm{kHz}$ from the pump laser, and the two output ports are detected with single-photon counting modules. The difference between the two photon-count streams yields a heterodyne beat note containing the cavity phase information.

Photon arrival events are binned with a time resolution of $2~\mu\mathrm{s}$. The resulting time trace is divided into $40~\mu\mathrm{s}$ segments, corresponding to a frequency resolution of 25 kHz and a sampling rate of 500 kHz. For each segment, we perform a Fourier transform and extract the phase from the complex Fourier component at the heterodyne frequency. The typical measurement data is shown in Fig.~\ref{fig:S2}b. The deterministic phase evolution of the LO is then subtracted to obtain the cavity phase.  For long time phase evolution measurements, the phase difference between adjacent time steps is mapped onto the principal interval \((-\pi,\pi]\) and then cumulatively summed to reconstruct a continuous unwrapped phase trajectory. The typical thermal drift of the optical path is $6~\mathrm{mrad}$ over 1 ms, which is negligible on the timescale of the measurement measurement.

\subsection{$U(1)$ Superradiance phase transition}
We derive the effective Hamiltonian for the system of atoms and the ring cavity that governs the $U(1)$-symmetry-breaking superradiance phase transition. We start from the Hamiltonian introduced in the main text,
\begin{equation} \label{eqs:Ham}
\begin{aligned}
\frac{\hat{H}}{\hbar}=&-\delta_s\left(\hat{a}_+^\dag\hat{a}_++\hat{a}_-^\dag\hat{a}_-\right)
+\frac{g^2}{\Delta}\left(\hat{a}_+\hat{a}_-^\dag\sum_{j=1}^Ne^{2ikx_j}+h.c.\right)\\&-\eta\left(i\hat{a}_+\sum_{j=1}^Ne^{ikx_j}+i\hat{a}_-\sum_{j=1}^Ne^{-ikx_j}+h.c.\right).
\end{aligned}
\end{equation}
In the dispersive regime, where the cavity shift induced by the atoms is negligible, $|\delta|\gg |Ng^2/\Delta|$, the second term on the RHS of Eq.(\ref{eqs:Ham}) corresponding to the cavity lattice with period $\lambda/2$ can be neglected. Within the mean-field approximation, the cavity-field operators are replaced by classical amplitudes $\alpha_\pm$. The equations of motion for the cavity fields $\alpha_\pm$ are
\begin{equation} \label{eqs:MFE}
\begin{aligned}
    &\dot{\alpha}_+=(i\delta_s-\frac{\kappa}{2})\alpha_++N\eta \langle S^*\rangle\\
    &\dot{\alpha}_-=(i\delta_s-\frac{\kappa}{2})\alpha_-+N\eta \langle S\rangle,
\end{aligned}
\end{equation}
where $\langle S\rangle $ is the mean value of the structure factor $S$ of atoms. For atoms at temperature $T$ in the lattice potential $U(x)=-\hbar\eta\left(i\alpha_+e^{ikx}+i\alpha_-e^{-ikx}+h.c.\right)$ given by the third term in the RHS of Eq.(\ref{eqs:Ham}), the structure factor
\begin{equation} \label{eqs:TES}
\begin{aligned}
\langle S\rangle=\langle e^{ikx}\rangle=\int_0^\lambda e^{ikx}p(x)dx,
\end{aligned}
\end{equation}
where $p(x)=\frac{e^{-U(x)/k_BT}}{\int_0^\lambda dxe^{-U(x)/k_BT}}$ is the Boltzmann distribution of atom position, expanding $p(x)$ to the third order in $\alpha_\pm$ yields
\begin{equation} \label{eqs:TES2}
\begin{aligned}
\langle S\rangle=&\frac{\hbar\eta}{k_BT}(-i\alpha_+^*+i\alpha_-)\\&-\frac{1}{2}\left(\frac{\hbar\eta}{k_BT}\right)^3|-i\alpha_+^*+i\alpha_-|^2(-i\alpha_+^*+i\alpha_-),
\end{aligned}
\end{equation}
Substitute Eq.(\ref{eqs:TES2}) into Eq.(\ref{eqs:MFE}), we can obtain
\begin{equation} \label{eqs:MFE2}
\begin{aligned}
    &\dot{\alpha}_+=(i\delta_s-\frac{\kappa}{2})\alpha_++\frac{N\hbar\eta^2}{k_BT}(i\alpha_+-i\alpha_-^*)\\
    &-\frac{1}{2}\frac{N\hbar^3\eta^4}{(k_BT)^3}|i\alpha_+-i\alpha_-^*|^2(i\alpha_+-i\alpha_-^*)\\
    &\dot{\alpha}_-=(i\delta_s-\frac{\kappa}{2})\alpha_-+\frac{N\hbar\eta^2}{k_BT}(-i\alpha_+^*+i\alpha_-)\\
    &-\frac{1}{2}\frac{N\hbar^3\eta^4}{(k_BT)^3}|-i\alpha_+^*+i\alpha_-|^2(-i\alpha_+^*+i\alpha_-).
\end{aligned}
\end{equation}
We perform a linear transformation and define the order parameter $\alpha=\frac{1}{\sqrt{2}}(-i\alpha_++i\alpha_-^*), \pi=\frac{1}{\sqrt{2}}(-i\alpha_+-i\alpha_-^*)$. Here, the order parameter $\alpha$ and $\pi$ are the optical-field analogues of position and momentum. We can rewrite Eq.(\ref{eqs:MFE2}) as the equations of motion for $\alpha$ and $\pi$:
\begin{equation} \label{eqs:MFE3}
\begin{aligned}
    &\dot{\alpha}=-\frac{\kappa}{2}\alpha+i\delta_s\pi\\
    &\dot{\pi}=-\frac{\kappa}{2}\pi+i(\delta_s+\frac{2N\hbar\eta^2}{k_BT})\alpha-i\frac{2N\hbar^3\eta^4}{(k_BT)^3}|\alpha|^2\alpha.
\end{aligned}
\end{equation}
From Eq.(\ref{eqs:MFE3}), we can obtain the effective Hamiltonian as 
\begin{equation} \label{eqs:MFH}
\begin{aligned}
\frac{H_{\text{eff}}}{\hbar}=-\delta_s|\pi|^2-(\delta_s+\frac{2N\hbar\eta^2}{k_BT})|\alpha|^2+\frac{N\hbar^3\eta^4}{(k_BT)^3}|\alpha|^4.
\end{aligned}
\end{equation}
The first term related to $\pi$ on the RHS of Eq.(\ref{eqs:MFH}) is the effective kinetic energy. The second and third terms of $\alpha$ are the potential energy. The potential of $\alpha$ takes the Landau form
\begin{equation} \label{eqs:MFV}
\begin{aligned}
V/\hbar=-A|\alpha|^2+B|\alpha|^4,
\end{aligned}
\end{equation}
with $A=-(\delta_s+\frac{2N\hbar\eta^2}{k_BT}),B=\frac{N\hbar^3\eta^4}{(k_BT)^3}.$
The phase transition occurs  when the quadratic coefficient $A=0$, corresponding to a critical threshold $\eta_c=\sqrt{\frac{k_B T}{2\hbar N}(-\delta_s)}$. When including the cavity decay $\kappa$, by solving the steady-state solutions of Eq.(\ref{eqs:MFE3}) for non-zero $\alpha$, the critical threshold $\eta_c$ is modified to
\begin{equation}
\eta_c=\sqrt{\frac{k_B T}{8\hbar N}\frac{4\delta_s^2+\kappa^2}{-\delta_s}}.  
\end{equation}
Above the threshold $\eta_c$, the system selects a non-zero value of $\alpha$, as the atoms self-organize and self-trap in the superradiant lattice and spontaneously break the $U(1)$ symmetry. This transition occurs when the potential energy gained from the self-organization overcomes the thermal energy that favors a disordered state. The excitation modes can be analyzed by solving the coupled equation of motion for $\alpha$ and $\pi$ as given in Eq.(\ref{eqs:MFE3}). 

\subsection{Collective motion of atoms and cavity fields}

In the self-organized phase, an emergent cavity lattice forms, and the atoms become tightly bound within it. However, because the ring cavity possesses $U(1)$ symmetry, the cavity lattice is not fixed in space but is instead translationally invariant. The atoms and the lattice can translate collectively along the cavity axis without a restoring force. The driving force for this motion arises from fundamental photon scattering of the lattice, which imparts random momentum kicks to the atoms, acting as a Langevin force. Since the cavity fields are generated by atoms and determined by their instantaneous positions, any atomic displacement caused by the Langevin force regenerates the cavity fields. The cavity fields, in return, exert back-actions on the atoms--on the one hand, it provides the trapping potential that binds them; on the other hand, its finite response time produces a friction force through the same retardation mechanism as cavity cooling. This interplay gives rise to a strongly interacting system of many atoms and the cavity fields. Under the combined action of the stochastic force and friction, the collective atom–lattice undergoes diffusion. We obtain the coupled equations of motion for atomic position and the cavity fields from Eq.~(\ref{eqs:Ham}) as
\begin{subequations}
\label{eqs:Eom1}
\begin{align}
\dot{\alpha}_+
&=
\left(i\delta_s-\frac{\kappa}{2}\right)\alpha_+
-i\frac{g^2}{\Delta}
\sum_{j=1}^{N}e^{-2ikx_j}\alpha_-
+\eta\sum_{j=1}^{N}e^{-ikx_j},
\label{eqs:Eom1a}
\\
\dot{\alpha}_-
&=
\left(i\delta_s-\frac{\kappa}{2}\right)\alpha_-
-i\frac{g^2}{\Delta}
\sum_{j=1}^{N}e^{2ikx_j}\alpha_+
+\eta\sum_{j=1}^{N}e^{ikx_j},
\label{eqs:Eom1b}
\\
m\ddot{x}_j
&=
-\hbar k
\left[
\frac{2ig^2}{\Delta}
\alpha_+\alpha_-^*e^{2ikx_j}
+\eta\left(\alpha_+-\alpha_-^*\right)e^{ikx_j}
+h.c.
\right] \nonumber\\
&~+f_j(t).
\label{eqs:Eom1c}
\end{align}
\end{subequations}
where $f_j(t)$ is the Langevin force acting on the $j$-th atom. 

To solve the collective motion from Eq.~(\ref{eqs:Eom1}), we note that in the self-organized phase, the atomic positions $x_j$ can be written as 
\begin{equation} \label{eqs:decomx}
\begin{aligned}
x_j(t)&=X(t)+m_j\lambda+\delta x^{\text{(osc)}}_j(t)+\delta x^{\text{(slow)}}_j(t).
\end{aligned}
\end{equation}
Here $X$ is the center-of-mass coordinate, and the position fluctuation around the perfect crystal position $X(t)+m_j\lambda$ is decomposed into two contributions. The first one is the fast oscillation $\delta x^{\text{(osc)}}_j(t)$ inside the superradiant lattice. This term changes at the lattice trapping frequency of typical $70~\mathrm{kHz}$, which is larger than the cavity linewidth $\kappa/(2\pi)=33.6~\mathrm{kHz}$, therefore the cavity field response to the fast oscillating motion is filtered out. The second one is the slow fluctuation $\delta x^{\text{(slow)}}_j(t)$ that changes on the same time scale as the center-of-mass motion  $X(t)$. In the low-temperature limit where the atomic temperature is less than the cavity lattice depth $k_\text{B} T < U$, the spatial extent of the atomic thermal distribution is much smaller than the lattice wavelength, so atoms are tightly confined to the lattice sites, moving as a rigid body. Both terms $\delta x^{\text{(osc)}}_j(t)$ and $\delta x^{\text{(slow)}}_j(t)$ are smaller than the lattice spacing $\lambda$, and much smaller than the overall drift $X$. The individual displacement of each atom is almost the same as the collective one, $\Delta x_j=\Delta X$. To describe the cavity field response to the atomic motion, we can expand the cavity fields to first order in the collective velocity, 
\begin{equation} \label{eqs:decomalpha}
\begin{aligned}
\alpha_\pm(t)&=\alpha_\pm+\dot{X}\alpha_\pm^{(1)},
\end{aligned}
\end{equation}
where $\alpha_\pm$ are the steady-state cavity fields for a stationary lattice and $\alpha_\pm^{(1)}$ is the correction coefficient for the cavity fields when atoms are moving. Substituting Eq.~(\ref{eqs:decomalpha}) into Eq.~(\ref{eqs:Eom1c}) and summing over all atoms, we obtain the equation of motion for the center-of-mass coordinate $X$ as
\begin{equation}
\label{eqs:Eom2c}
\begin{aligned}
M\ddot{X}
=&-N\hbar k
\Bigg\{
\frac{2ig^2}{\Delta}
\left(
\alpha_+^{(1)}\alpha_-^{*}
+\alpha_+\alpha_-^{(1)*}
\right)
e^{2ikX}
\\
&\quad
+\eta
\left(
\alpha_+^{(1)}
-\alpha_-^{(1)*}
\right)
e^{ikX}
+h.c.
\Bigg\}\dot{X}
+F(t),
\end{aligned}
\end{equation}
where $M=Nm$ is the total atomic mass, and $F(t)=\sum_j f_j(t)$. In deriving Eq.~(\ref{eqs:Eom2c}), we use the fact that all atoms at the equilibrium positions $x_j=m_j\lambda$ are located at the minima of the superradiant lattice, where the conservative force vanishes. The force arises from the cavity's delayed response to the atomic displacement from the instantaneous lattice potential minima, resulting in a damping force. Each atom experiences a force, $-\beta \dot X$, that is proportional to the collective velocity $\dot X$, with the damping coefficient $\beta$ given by the product of the cavity steady-state fields $\alpha_\pm$, and the motion-induced first-order correction coefficients $\alpha_\pm^{(1)}$, as
\begin{equation}
\begin{aligned}
\beta=\hbar k\bigg[&\frac{2ig^2}{\Delta}\left(\alpha_+^{(1)}\alpha_-^{*}+\alpha_+\alpha_-^{(1)*}\right)e^{2ikX}\\&+\eta\left(\alpha_+^{(1)}-\alpha_-^{(1)*}\right)e^{ikX}+h.c.\bigg].
\end{aligned}
\label{eqs:gamma1}
\end{equation}

To find the explicit solutions for $\alpha_\pm$ and $\alpha_\pm^{(1)}$, we substituting Eq.~(\ref{eqs:decomalpha}) into Eq.~(\ref{eqs:Eom1a}) and Eq.~(\ref{eqs:Eom1b}) and solve for the cavity fields in the zeroth and first order in $\dot X$ respectively. We obtain

\begin{subequations}
\begin{equation}
\begin{aligned}
&0=(i\delta_s-\frac{\kappa}{2})\alpha_+-i\frac{Ng^2}{\Delta}e^{-2ikX}\alpha_-+N\eta e^{-ikX}\\
&0=(i\delta_s-\frac{\kappa}{2})\alpha_--i\frac{Ng^2}{\Delta}e^{2ikX}\alpha_++N\eta e^{ikX},
\end{aligned}
\label{eqs:Eom2a}
\end{equation}
\begin{equation}
\begin{aligned}
&\frac{\partial\alpha_+}{\partial X}=(i\delta_s-\frac{\kappa}{2})\alpha_+^{(1)}-i\frac{Ng^2}{\Delta}e^{-2ikX}\alpha_-^{(1)}\\
&\frac{\partial\alpha_-}{\partial X}=(i\delta_s-\frac{\kappa}{2})\alpha_-^{(1)}-i\frac{Ng^2}{\Delta}e^{2ikX}\alpha_+^{(1)}.
\end{aligned}
\label{eqs:Eom2b}
\end{equation}

\end{subequations}

Therefore,
\begin{align}
\alpha_\pm
&=
\frac{e^{\mp i k X}N\eta}
{-i\left(\delta-\frac{2Ng^2}{\Delta}\right)+\kappa/2},
\label{eqs:sss}
\\[4pt]
\alpha_\pm^{(1)}
&=
\frac{\mp i k\alpha_{\pm}}
{i\delta-\kappa/2}.
\label{eqs:fss}
\end{align}

Substituting Eq.~(\ref{eqs:sss}) and ~(\ref{eqs:fss}) into Eq~(\ref{eqs:gamma1}), we obtain
\begin{equation}
\label{eqs:gammaFull}
\begin{aligned}
\beta
=&\,
k^2
\Bigg[
-\frac{16\hbar g^2}{\Delta}
\frac{\kappa}{4\delta^2+\kappa^2}
|\alpha_\pm|^2
\\
&\quad
+
U
\frac{2\kappa}{4\delta^2+\kappa^2}
\left(
1+
\frac{\delta}
{\delta-2Ng^2/\Delta}
\right)
\Bigg],
\end{aligned}
\end{equation}
where $U = 4\hbar\eta|\alpha_\pm|\sin{\Phi}$ is the trap depth of the cavity lattice. The second term on the RHS of Eq.~(\ref{eqs:gammaFull}) corresponds to the usual ring-cavity cooling contribution and becomes dominant in the limit $|\delta|\gg |Ng^2/\Delta|$. For the experimental parameter range, particularly for $N\gtrsim30$, the collective dispersive shift is not always negligible. 
We therefore retain the full expression in Eq.~(\ref{eqs:gammaFull}) in the calculation of the damping coefficient in Fig.~4. 

The equation of motion for the collective coordinate $X$ becomes
\begin{equation} \label{eqs:Eom3}
\begin{aligned}
M\ddot{X}=-N\beta\dot{X}+ F(t).
\end{aligned}
\end{equation}
Here, the total Langevin force acting on the collective coordinate $X$ is $F(t)=\sum_j f_j(t)$, imparting random momentum kicks to the rigid atom–lattice body. The total random force $F(t)$ satisfies the white-noise correlation $\langle F(t)F(t')\rangle=S_{FF}\delta(t-t')=N\hbar^2k^2\Gamma_{\text{sc}}\delta (t-t')$, where $\Gamma_{\text{sc}}$ is the photon scattering rate for a single atom. For the ring cavity, the single-atom scattering rate is $\Gamma_{\text{sc}}=(\frac{3}{10}+\frac{2C\kappa^2}{4\delta^2+\kappa^2})\Gamma_{\text{fs}}$, where $\Gamma_{\text{fs}}$ is the free-space scattering rate of a pumped atom inside the cavity. The first term describes the absorption of a pump photon followed by spontaneous emission into free space. For a circularly polarized pump, the factor $3/10$ arises from the angular projection of the dipole-radiation pattern onto the cavity axis. The second term describes pump-photon scattering into the cavity modes and is enhanced by the single-atom cooperativity $C$, with the detuning dependence determined by the Lorentzian cavity response.

Equation~(\ref{eqs:Eom3}) shows that the collective coordinate $X$ undergoes diffusion. Beyond the relaxation time such that $t\gg M/(N\beta)$, the variance of the center-of-mass coordinate is
\begin{equation} \label{eqs:diffution}
\begin{aligned}
\Delta X^2=\frac{N\hbar^2 k^2 \Gamma_{\text{sc}}}{\left(N\beta\right)^2} t=\frac{\lambda^2\Gamma_{\text{sc}}}{N\tilde{\beta}^2}t=2\mathcal{D}t.
\end{aligned}
\end{equation}
where $\mathcal{D}$ is the diffusion constant and $\tilde{\beta}=2\pi\beta/(\hbar k^2)$ is the normalized damping coefficient. 

\subsection{Diffusion at finite temperature}
The analytical result for the diffusion constant $\mathcal{D}$ in Eq.~(\ref{eqs:diffution}) is valid in the $T=0$ limit, where atoms are perfectly pinned by the dynamical lattice with the structure factor $S=1$. At finite temperature, the thermal motion melt the atomic crystal, leading to a correction to the diffusion constant. To quantify these effects, we numerically solve the coupled equations for the cavity fields and atomic motion, Eq.~(\ref{eqs:Eom1}). 

The simulation follows the actual experiment sequence, in which the pump beam is gradually ramped up while the optical tweezer traps are switched off, thereby adiabatically loading the atoms into the optical lattice. The initial atomic positions and velocities are sampled from thermal Boltzmann distributions in tweezers. The stochastic force on the $j$-th atom is written as $f_j=\sqrt{\hbar^2k^2\Gamma_{\text{sc}}}\xi_j(t)$, where $\xi_j(t)$ is a normalized white-noise process satisfying $\langle \xi_j(t)\xi_l(t')\rangle=\delta_{jl}\delta(t-t')$. The white noise is approximated by a piecewise constant process over each integration time step $\xi_j(t)\approx \frac{r_{j,n}}{\sqrt{\Delta t}}$ for $t\in [(n-1)\Delta t,n\Delta t]$, where $r_{j,n}\sim \mathcal{N}(0,1)$ are independent Gaussian random variables for different atoms and different time steps. 

\begin{figure}[htb]
\includegraphics[width=9cm,angle=0]{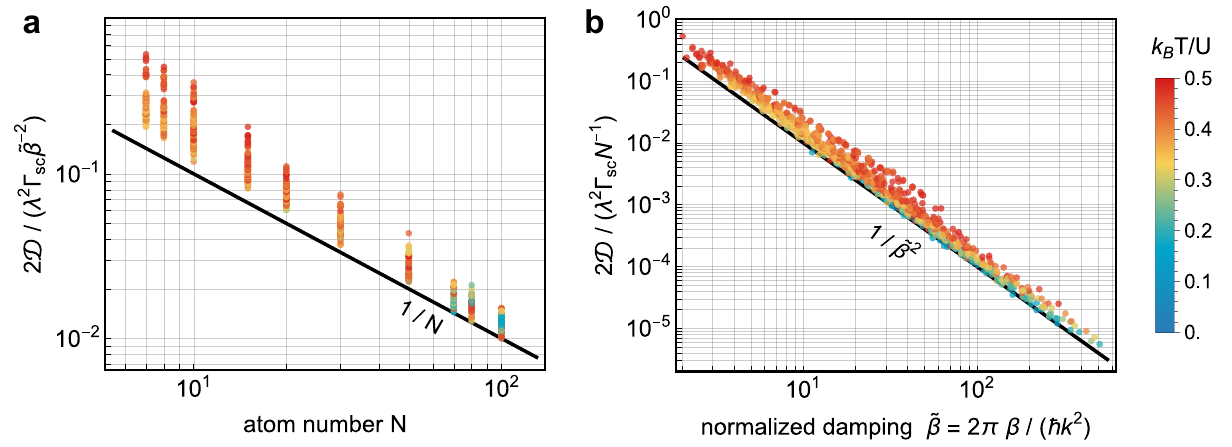}
\caption{\label{fig:S3}\textbf{Collective diffusion at finite temperature.} \textbf{a,} Normalized diffusion rate $2\mathcal{D}/\left( \lambda^2 \Gamma_{\text{sc}}\tilde{\beta}^{-2}\right)$ as a function of atom number $N$. Each point is obtained from numerical simulations performed under different parameter sets spanning the experimental range, including variations of pump strength, pump–cavity detuning, and atom number. The dashed line indicates the zero-temperature prediction $\mathcal{D}\propto1/N$.
\textbf{b,} Normalized diffusion rate $2\mathcal{D}/\left( \lambda^2 \Gamma_{\text{sc}}N^{-1}\right)$ as a function of the normalized damping coefficient $\tilde{\beta}=2\pi\beta/(\hbar k^2)$. The dashed line shows the zero-temperature scaling $\mathcal{D}\propto1/\tilde{\beta}^2$. In \textbf{a} and \textbf{b}, the color of each point denotes the ratio $k_B T/U$. For lower temperatures, the simulations closely follow the universal analytical scaling laws, whereas stronger thermal fluctuations lead to clear deviations.}
\end{figure}

To reveal the universality of the diffusion constant at finite temperature, we extract the diffusion constant and the ratio $k_B T/U$ from simulations over the experimentally accessible parameter range, varying the pump Rabi frequency $\Omega$, pump-cavity detuning $\delta$, atom number $N$, and the atomic temperature. In Fig.~4a,b, we plot the normalized diffusion constant as a function of the atom number $N$ and normalized damping coefficient $\tilde{\beta}$, with the shaded bands corresponding to the experimental temperature range the experimental range $0.3<k_BT/U<0.5$. Fig.~\ref{fig:S3}a,b extend the simulations to a range $0<k_BT/U<0.5$, greater than the experimental temperature range. At lower temperature, the simulated data agree better with the analytical prediction at zero temperature and exhibit the universal $1/N$ and $1/\tilde{\beta}^2$ scalings, while higher temperature causes deviations. From the simulations, we also extract the atomic structure factor amplitude $|S|$ as a function of $k_BT/U$, as shown in Fig.~4c.

\subsection{Long-time breakdown of diffusion and melting of the atomic crystal}
\begin{figure}[htbp]
\includegraphics[width=7.5cm,angle=0]{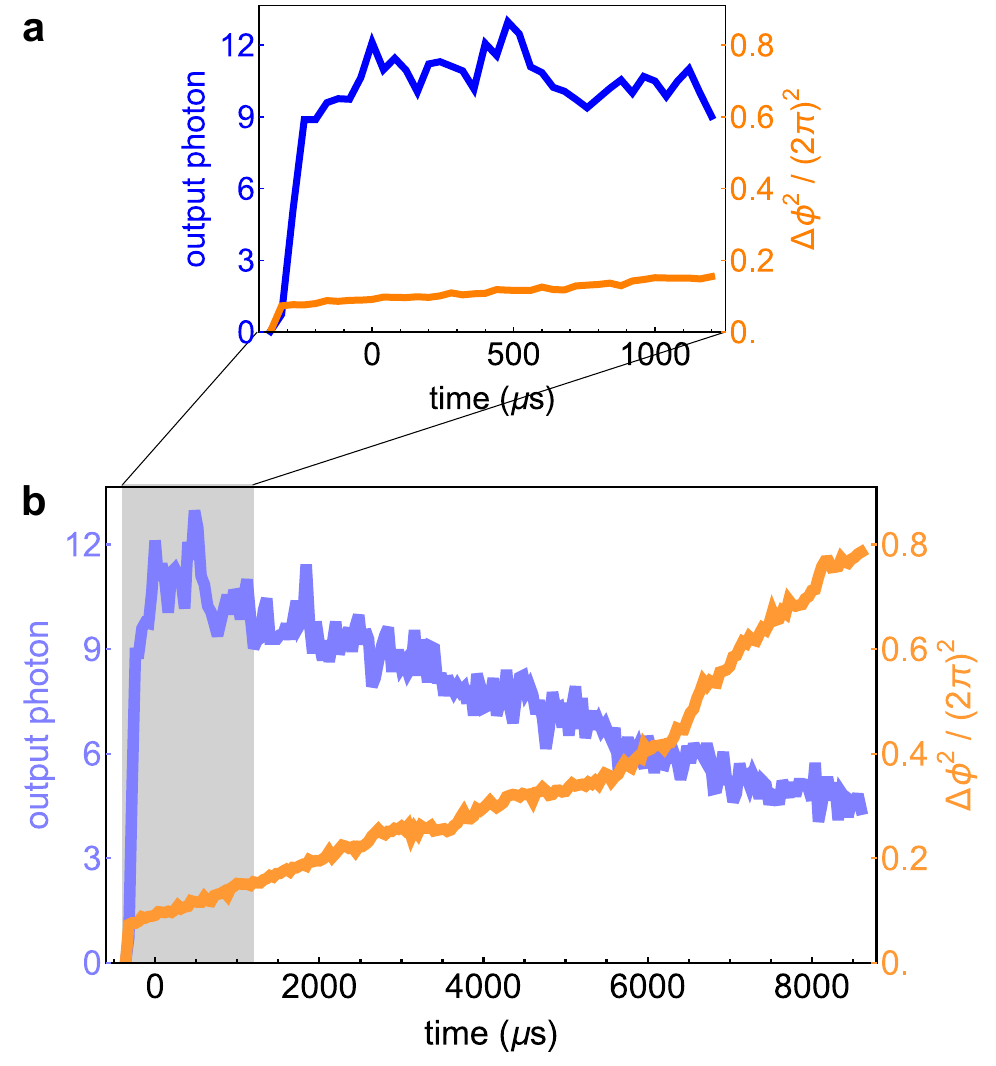}
\caption{\label{fig:S4}\textbf{Long-time evolution and technical heating of the self-organized state.} \textbf{a,} Short-time dynamics of the output photon number and phase variance after preparation of the self-organized state. The cavity transmission rapidly reaches a steady value, while the phase variance grows slowly. \textbf{b,} Extended evolution over several milliseconds. The output photon number gradually decays, indicating heating-induced loss of atoms from the superradiant lattice, while the phase variance exhibits nonlinear growth as the collective atom–lattice motion becomes increasingly affected by heating. The shaded region corresponds to the time window shown in \textbf{a}.}
\end{figure}

In the main text, we show the diffusion of self-organized atoms self-trapped in the emergent cavity superradiant lattice. Over our observation time, the cavity output power remains constant, and the phase variance increases linearly. At longer evolution times, however, we observe a gradual decline in cavity transmission along with nonlinear growth of the phase variance. This occurs because the atoms are gradually heated and the self-organized lattice begins to melt, causing the superradiant lattice depth to drop. The system dynamics deviates from the collective diffusion regime. The heating arises from two sources: (i) the fundamental spontaneous scattering from the cavity lattice, and (ii) parametric heating of the cavity lattice due to conversion of residual cavity-lock frequency noise into lattice intensity noise. When the laser lock to the cavity is not optimized, we observe significantly faster parametric heating that obscures the correct diffusion behavior. With an optimized laser-cavity lock, we obtain a sufficiently long time window for the atomic crystal to sustain itself. An example of a long-time measurement is shown in Fig.~\ref{fig:S4}. All diffusion constants reported in the main text are extracted well before this heating-induced decay becomes significant.

\subsection{Ring cavity back scattering}
A gapless Goldstone mode requires continuous $U(1)$ symmetry and is inherently fragile against symmetry-breaking perturbations. Our ring cavity provides a platform with near-perfect $U(1)$ symmetry, despite a known technical imperfection: residual backscattering from the cavity mirrors can, in principle, break the ideal symmetry. In our system, however, the backscattered power is only 0.5\%, producing a parasitic optical lattice with a depth of approximately 2\% of the superradiant lattice depth. This residual lattice is also well below the atomic thermal energy. As a result, the evolution of the system within the $U(1)$ manifold remains almost unaffected by this small mirror-induced lattice. The experimentally measured scaling $\mathcal{D} \propto 1/( N \tilde{\beta}^2)$ confirms that this minor technical imperfection is negligible and does not compromise the Goldstone mode.

\end{document}